\documentclass[preprint,prd,aps,tighten,nofootinbib,amssymb]{revtex4}
\usepackage{color}
\input{colordvi.tex}
\usepackage{amsmath}
\usepackage[dvips]{graphicx}
\usepackage{subfigure}
\usepackage{comment}
\newcommand{\beq}{\begin{equation}}
\newcommand{\eeq}{\end{equation}}
\newcommand{\bes}{\begin{split}}
\newcommand{\ees}{\end{split}}
\newcommand{\beqa}{\begin{eqnarray}}
\newcommand{\eeqa}{\end{eqnarray}}

\newcommand{\bea}{\begin{eqnarray}}
\newcommand{\eea}{\end{eqnarray}}
\newcommand{\bear}{\begin{array}}
\newcommand {\eear}{\end{array}}
\newcommand{\bef}{\begin{figure}}
\newcommand {\eef}{\end{figure}}
\newcommand{\bec}{\begin{center}}
\newcommand {\eec}{\end{center}}

\newcommand{\la}{\left\langle}
\newcommand{\ra}{\right\rangle}

\def\GEV#1{10^{#1}{\rm\,GeV}}

\def\lrf#1#2{ \left(\frac{#1}{#2}\right)}

\begin{document}
\widetext
\draft

\begin{flushright}
KEK-TH-1643\\
UT-13-25\\
TU-940\\
IPMU13-0133
\end{flushright}

\title{
Cosmological constraints on axionic dark radiation from\\ axion-photon conversion in the early Universe
}

\author{Tetsutaro Higaki$^a$, Kazunori Nakayama$^{b,d}$ and Fuminobu Takahashi$^{c,d}$}

\affiliation{$^a$Theory Center, KEK, 1-1 Oho, Tsukuba, Ibaraki 305-0801, Japan \\
$^b$Department of Physics, The University of Tokyo, Tokyo 113-0033, Japan \\
$^c$Department of Physics, Tohoku University, Sendai 980-8578, Japan\\
$^d$Kavli Institute for the Physics and Mathematics of the Universe (WPI), Todai Institutes for Advanced Study, The University of Tokyo, Kashiwa 277-8583, Japan}

\date{\today}

\begin{abstract}
Axions seem ubiquitous in string theories and some of them may be produced non-thermally by heavy scalar decays, 
contributing to dark radiation.  
We study various cosmological effects of photons produced from the axionic dark radiation 
through axion-photon conversion  in  the presence of primordial magnetic fields, and derive 
tight constraints on the combination of the axion-photon coupling and the primordial
magnetic field.
\end{abstract}

\maketitle

\section{Introduction}
\label{sec:intro}

String theory is a promising candidate for the unified theory.
After the compactification of string theory, there often appear light moduli,
whose mass scales are of order of the supersymmetry (SUSY) breaking scale or even lighter.
Such light moduli tend to dominate the energy density of the Universe and cause various 
cosmological problems, known as the cosmological moduli problem~\cite{Coughlan:1983ci,de Carlos:1993jw}.
This is one of the most important constraints on building realistic string-motivated particle physics models.

Among many solutions proposed so far~\cite{Linde:1996cx,Dine:1998qr,Randall:1994fr,Lyth:1995ka,Kawasaki:2004rx,
Nakayama:2011wqa}, the simplest one is to make the moduli heavy enough to decay before the big-bang nucleosynthesis (BBN) begins.
In this case, one has to make sure if the moduli decay do not produce cosmologically dangerous relics. In fact, it is known that
the moduli generically decay into gravitinos with a sizable branching fraction, if kinematically allowed~\cite{Endo:2006zj}.\footnote{
The decay of thus produced gravitinos may spoil BBN, or produce
too many lightest SUSY particles. The moduli-induced gravitno problem can be solved if the gravitino as well as the lightest SUSY particle
is heavier than the moduli or the R-parity is broken by a small amount.}
More importantly, the present authors recently pointed out a generic problem that appears in many scenarios where the lightest moduli fields are 
stabilized  by SUSY  breaking effects:  those moduli fields  tend to mainly decay into almost massless axions, whose abundance is tightly 
constrained by the recent Planck results~\cite{Higaki:2013lra}.\footnote{The presence of extremely light axion is ensured by the shift symmetry : $T\to T + i \beta$,
where $T$ denotes the modulus field and $\beta$ is a real constant.
Then ${\rm Im}\,T$ is identified as an axion, which obtains a mass only through non-perturbative effects.}  The moduli-induced axion problem 
 cannot be solved simply by increasing the modulus mass, and so, it is a rather robust problem. 
Not only does it  place stringent limits on the moduli stabilization mechanism and the modulus couplings  with
the visible sector, but it also suggests that the axionic dark radiation may be ubiquitous in string theories. 
The axion production from the modulus decay was already known and studied extensively based on concrete examples. See e.g. Refs.~\cite{Higaki:2012ba,Cicoli:2012aq,
Higaki:2012ar,Angus:2013zfa}. 
%
%

Without invoking the string theoretic setup, it is a generic feature that the saxion in SUSY QCD axion models often decays into a pair of 
axions~\cite{Chun:1995hc, Higaki:2011bz}.
The possibility of axionic dark radiation in SUSY axion models was studied in many literatures~\cite{Choi:1996vz,Hashimoto:1998ua,Chun:2000jr,Ichikawa:2007jv,Kawasaki:2011ym,Kawasaki:2011rc,Jeong:2012hp,Moroi:2012vu,Choi:2012zna,Graf:2012hb}.
In this case, the QCD axion may account for both dark matter and dark radiation. 
Thus, the presence of axionic dark radiation is highly motivated by theoretical models beyond the standard model. 

The Planck results constrained the amount of dark radiation as
$N_{\rm eff} = 3.30^{+0.54}_{-0.51}$\,$(95\%$\,C.L.)
in terms of the effective number of neutrino species~\cite{Ade:2013lta}.
Interestingly, the observations give a slight preference to the existence of dark radiation, 
$\Delta N_{\rm eff} \equiv N_{\rm eff} - 3.046 > 0$.
Therefore, the dark radiation may provide 
a clue to physics beyond the standard model. In order to distinguish between various
dark radiation models, one needs the novel methods to detect dark radiation
through their interactions.

In this paper we study cosmological effects of axionic dark radiation under the presence of primordial magnetic field.
It is well known that axions are converted into photons and vice versa in the magnetic field background
if the axion has an interaction with the electromagnetic field of the form~\cite{Sikivie:1983ip,Raffelt:1987im}\footnote{
There are on-going and planned experiments for  axion-like particle
 search~\cite{Asztalos:2009yp,FerrerRibas:2012ru,Ehret:2009sq,Vogel:2013bta,Bahre:2013ywa,Horns:2012jf}.
The cosmological and terrestrial constraints on $g_a$ are summarized in Ref.~\cite{Cadamuro:2011fd}. }
\begin{equation}
	\mathcal L = -\frac{1}{4}g_a a F_{\mu\nu}\tilde F^{\mu\nu},
\end{equation}
where $a$ denotes the axion field, $g_a$ is the coupling constant,
$F_{\mu\nu}$ is the field strength of the electromagnetic gauge field and $\tilde F^{\mu\nu} = \epsilon^{\mu\nu\rho\sigma}F_{\rho\sigma}/2$.
If relativistic axions are converted into photons in the early Universe, 
they may cause disastrous effects on BBN, cosmic microwave background (CMB), etc.
Hence the presence of axionic dark radiation may already be constrained from observations.
Although there are no consensus on the strength of the intergalactic magnetic field, the recent Fermi data  indicate lower bounds
on the intergalactic magnetic field of the order $B_0 \gtrsim 10^{-15}$\,G on the scale of $\ell_B \gtrsim 1$\,Mpc,
and it can be as large as $\sim 1$\,nG~\cite{Neronov:1900zz}.
If these magnetic fields have a primordial origin, they necessarily cause axion-photon mixing in the early Universe.
	For the discussion on the origin of primordial magnetic field, see Refs.~\cite{Grasso:2000wj,Kandus:2010nw,Durrer:2013pga}.

In the case of the string-theoretic axions,
it has been discussed that axions whose decay constants are given by $f_{a}\sim \frac{1}{8\pi^2g_a}\sim 10^{14}$ GeV 
for the Calabi-Yau volume ${\cal V} \sim 10^7$ in Refs.~\cite{Conlon:2006tq,Cicoli:2012sz} 
(see also \cite{Higaki:2011me} for general discussions on the axiverse).
The axions are coupled to both QCD and $U(1)_{\rm EM}$,
when the visible gauge couplings are given by blowing-up local moduli which 
are stabilized by string-loop corrections. Such axions could acquire (ultralight) masses via non-perturbative effects. 
It is expected that
axionic dark radiation is produced through the lightest modulus decay because
the axions are coupled to the lightest modulus in their kinetic terms~\cite{Cicoli:2012aq,Higaki:2012ar}.\footnote{
In this case, we expect large soft masses as the gravitino mass and hence 
one needs large coefficients in the Giudice-Masiero terms or many light Higgs fields through a fine-tuning.
}

Lastly let us mention the related works in the past. 
The conversion of CMB into the axion in the primordial magnetic field and resulting constraints were studied in Refs.~\cite{Yanagida:1987nf,Mirizzi:2009nq}.
Some cosmological effects of axionic dark radiation were studied in Refs.~\cite{Conlon:2013isa,Conlon:2013txa}.
Ref.~\cite{Conlon:2013isa} considered scatterings of the relativistic axions with matter, and studied the
BBN constraint as well as SUSY particle production.
More recently, Ref.~\cite{Conlon:2013txa} studied the axion conversion into the X-ray photons in the cluster magnetic field.
In this paper, we study the axion-photon conversion under the primordial magnetic field,
instead of the axion scattering or the cluster magnetic field. As a result, we derive
tight constraints on  combination of the axion-photon coupling $g_a$ and the primordial
magnetic field $B$ for a wide range of the axion mass.
As we shall see later, it can even exclude the QCD axion as substantial dark radiation for some parameters.

In Sec.~\ref{sec:conv} we formulate the method to calculate the axion-photon conversion probability,
including the resonant conversion.
In Sec.~\ref{sec:const} we derive cosmological constraints on the axionic dark radiation.
Sec.~\ref{sec:disc} is devoted to  discussion and conclusions.

\section{Axion-photon conversion in the early Universe}
\label{sec:conv}

The purpose of this section is to evaluate the conversion probability of the ultra-relativistic 
axions into photons through the mixing  induced by the background magnetic field.
The energy of axions at present is denoted by $E_0$, which is assumed to be much higher than the 
temperature of the CMB photons, $T_0 \simeq 2.725$\,K, i.e., $E_0 \gg T_0$.
This enables us to start from the initial condition being the pure axion state, as
there is effectively no background photons with such high energy.  Throughout this paper,
we assume that the axion is relativistic until present, namely, $E_0 \gg m_a$, where $m_a$
is the axion mass. In the early Universe the axion energy scales as $E = E_0 (1+z)$, where $z$ is the redshift parameter.

Let us start with the following Lagrangian 
\begin{equation}
	\mathcal L = \frac{1}{2}(\partial a)^2-\frac{1}{4}F_{\mu\nu}F^{\mu\nu} - \frac{1}{2}m_a^2a^2 
	- \frac{1}{4}g_a a F_{\mu\nu}\tilde F^{\mu\nu},
	\label{L}
\end{equation}
where $F_{\mu\nu}$ is the field strength of the electromagnetic gauge field and $\tilde F^{\mu\nu} = \epsilon^{\mu\nu\rho\sigma}F_{\rho\sigma}/2$.
Under the background magnetic field $\vec B$, the last term in (\ref{L}) induces the axion-photon mixing~\cite{Raffelt:1987im}.
Our results do not depend on the sign of $g_a$.

In analogy with the neutrino oscillation, the axion-photon oscillation can be described in terms of the density matrix~\cite{Stodolsky:1986dx,Sigl:1992fn}:
\bea
(2\pi)^3 \delta^{(3)}({\bf p} - {\bf q}) \left[\rho_{\bf p}\right]_{ij} &\equiv & \la a_j^\dag({\bf p}) a_i({\bf q}) \ra.
\eea
where $a_i({\bf p})$ and $a_i^\dag({\bf p})$ denote the annihilation and creation operators of the $i$-th particle
with three-momentum ${\bf p}$, respectively. The density matrix is a generalized version of the occupation number,
and it is given by a $3 \times 3$ matrix since photons have two polarization states.  
For a given constant  ${\vec B}$, however, it is one of the two polarization states that gets mixed with axions. Therefore,
the density matrix is  represented by a $2 \times 2$ matrix   ($i=1$: photons, $i=2$: axions)
as long as one considers the axion conversion under the constant magnetic field.
The effect of passing  through regions with different ${\vec B}$ can be effectively taken into account in the two-flavor
regime, as we shall see shortly. Such simplification is sufficient for the order-of-magnitude estimate of the conversion 
probability of axions into photons.
%

In the case of two flavor oscillations, it is useful to expand the density matrix in terms of the Pauli matrices;
\begin{equation}
	\rho_{\bf p} \equiv \frac{1}{2}(P_0+\vec\sigma \cdot {\bf P}) 
	= \frac{1}{2}\begin{pmatrix}
		P_0+P_z & P_x - iP_y \\
		P_x+iP_y & P_0-P_z	
	\end{pmatrix},
\end{equation}
where $P_x$ and $P_y$ is the correlation between photons and axions, and $P_z (P_0)$ represents  difference (sum) of the photon and axion abundances.
If ${\bf P} = 0$, there is no correlation, and in particular, there is an equal amount of photons and axions. Such state is referred to as being 
in ``flavor equilibrium.''
We are interested in the conversion probability of axions into photons. For this purpose the overall normalization of the density matrix is
not relevant, and so,  we adopt the normalization of the density matrix such that 
$P_0=P_z = 1$ and $P_0=-P_z=1$ represent the pure photon and axion states, respectively. 
We use the pure axion state as the initial condition, and  follow the evolution of $\bf P$.
Then  $\rho_{{\bf P} 11}$ $(\rho_{{\bf P} 22})$ represents the probability that the photon (axion) is found.

The evolution of $P_0$ and $\bf P$ is described by the following equations~\cite{Stodolsky:1986dx,Sigl:1992fn}:\footnote{
	A similar system consisting of an active neutrino and a sterile neutrino is considered in e.g. Ref.~\cite{Hannestad:2012ky}.
}
\begin{equation}
\begin{split}
	&\frac{\partial}{\partial t}P_0  = - D_0 (P_0 + P_z),\\
	&\frac{\partial}{\partial t}{\bf P} = {\bf V} \times {\bf P} - D {\bf P}_T +\dot P_0  {\bf e_z},
	\label{Peq}
\end{split}
\end{equation}
where ${\bf P}_T = (P_x,P_y,0)$, ${\bf e_z} = (0,0,1)$ and
\begin{equation}
	{\bf V} = (V_x,V_y,V_z) = \left( g_a B, 0, \frac{\omega_p^2- m_a^2}{2E}  \right).
\end{equation}
Here $B$ represents the magnetic field transverse to the wave propagation direction.
The decoherence effect of photon scatterings  with the background plasma as well as passing through
the magnetic field domains is taken into account by adding damping terms with the coefficients $D_0$ and $D$, which are given by
 $D_0  \equiv \sigma_{\gamma e} n_e/2$ and $D \equiv D_0 + \ell_B^{-1}$ with
 $\sigma_{\gamma e} = \sigma_T$ for $E < m_e$ and $\sigma_{\gamma e} \sim \sigma_T (m_e/E)$ for $E > m_e$.
Here $\sigma_T$ is the Thomson scattering cross section, $n_e$ the electron number density,
$\ell_B$ the coherent length of the magnetic field, 
and $E=E_0(1+z)$ the axion energy.
The plasma frequency $\omega_p$ is given by
\begin{equation}
	\omega_p = \sqrt{\frac{4\pi \alpha n_e}{m_e}} \simeq 2\times 10^{-14}\,{\rm eV}\, (1+z)^{3/2} X_e^{1/2},
\end{equation}
where $\alpha$ is the fine structure constant, $m_e$ the electron mass and $n_e$ the electron number density.
For the photon with energy lower than the ionization energy of the hydrogen atom,
the ionized fraction $X_e$ is taken to be $X_e = 1$ for $z > 1090$ and $z < 11.4$ while $X_e\simeq 2\times 10^{-4}$ for
$11 < z < 1090$ as indicated by the Planck results.
On the other hand,   photons with energy higher than the ionization energy 
 do not distinguish free electrons from those bound in atoms. 
Thus we simply set $X_e=1$ for $E > 13.6$\,eV in the evaluation of the plasma frequency,
independently of the redshift. 
While the evolution of the magnetic field strength depends on the model of magnetogenesis,
we assume the simple scaling $B \simeq B_0 (1+z)^2$ as it is realized if the large scale magnetic field is generated in the early Universe (say, during inflation) and frozen into the medium~\cite{Grasso:2000wj}.

Let us comment on the effect of the coherent length of the magnetic field, $\ell_B$.
For simplicity we adopt the conventional cell model for the primordial magnetic field, 
in which the magnetic field is given by a constant vector in each cell,
and there is no correlation of the magnetic fields between the adjacent cells. 
We adopt the following value of the coherent length:
\beq
	\ell_B \sim 1\,{\rm Mpc}\,(1+z)^{-1} \simeq 1.6\times 10^{29}\,{\rm eV^{-1}} (1+z)^{-1}.
\eeq
The axion-photon mixing is interrupted and the correlation between photons and axions is suppressed
 each time  axions pass through the boundary of the cells. The sudden change of the background (classical) 
 magnetic field may be interpreted as the measurement of the quantum system of axions and photons, in analogy with 
 the Stern-Gerlach experiment.  This is the reason why we included $\ell_B^{-1}$ in the definition of $D$. 
 Precisely speaking, the polarization state of photons which mixes with axions depends on the direction of ${\vec B}$,
 and the evolution of the density matrix should be described in terms of a $3 \times 3$ matrix. However,
 the above simplistic treatment in the $2 \times 2$ matrix could be used for the order of magnitude estimate
 of the conversion rate of axions into photons. 
%
 Note that the photons with an energy $E$ disappear at the rate of $D_0$ and its energy dissipates into the plasma,
while the photons are still propagating along the same direction after   passing through the
cells of the magnetic field. That is why  only $D_0$ appears as the damping term in the evolution equation of $P_0 + P_z$. 

In the expanding Universe, the evolution equations (\ref{Peq}) are slightly modified, as the momentum as well as the energy
are redshifted. In effect, the evolution equations (\ref{Peq}) are valid in about one Hubble time. 
Instead of following the evolution of the axion-photon system all the way down to present from the axion production, 
we analytically estimate the conversion rate in one Hubble time during which the evolution equations hold approximately. 

\begin{figure}[t]
\begin{center}
\includegraphics[scale=1.2]{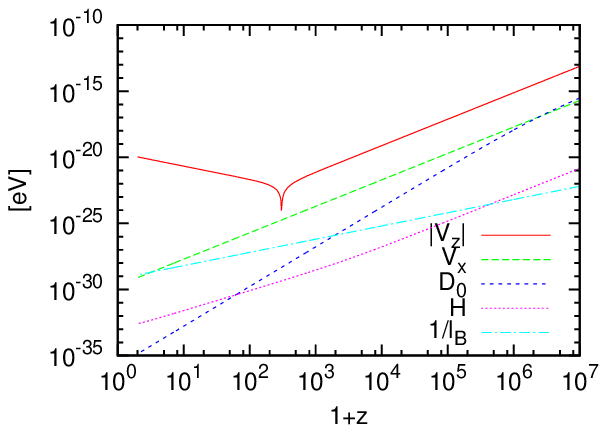} 
\includegraphics[scale=1.2]{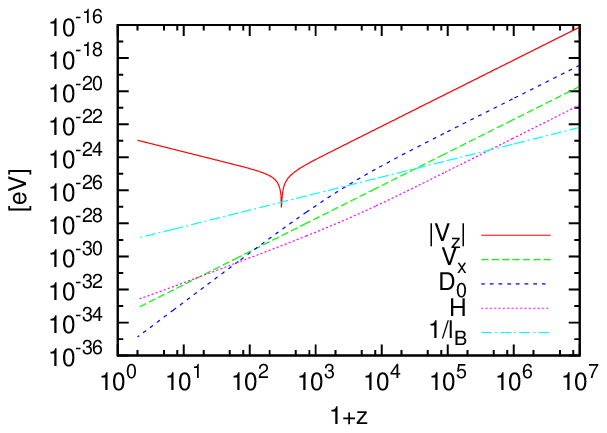} 
\caption{
	Various quantities as a function of redshift for $m_a = 10^{-10}$\,eV and
	$E_0/T_0 = 10^3$, $g_a B_0 = 10^{-10}\,{\rm GeV^{-1}nG}$
	(left) and $E_0/T_0 = 10^6$, $g_a B_0 = 10^{-14}\,{\rm GeV^{-1}nG}$ (right).
}
\label{fig:time}
\end{center}
\end{figure}

\begin{figure}[t]
\begin{center}
\includegraphics[scale=1.5]{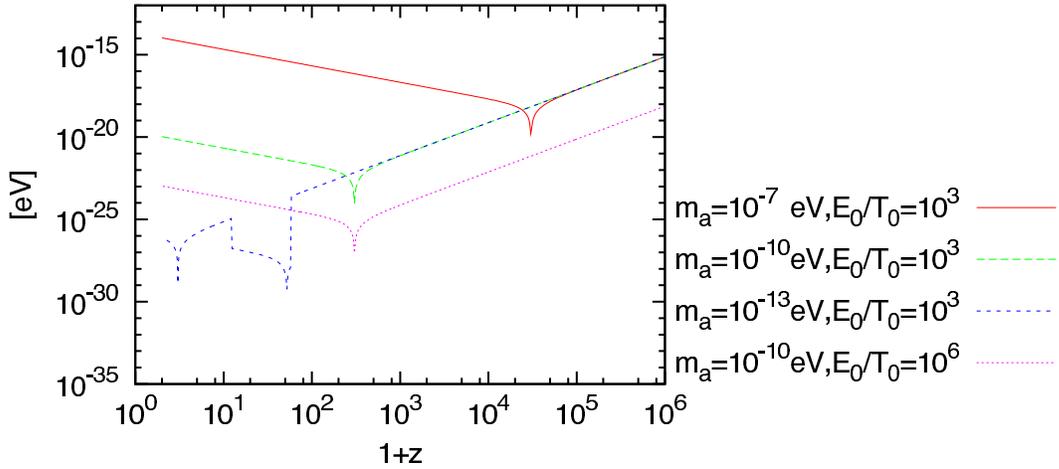} 
\caption{
	$|V_z|$ as a function of redshift.
}
\label{fig:Vz}
\end{center}
\end{figure}

Let us summarize here the redshift dependence of various quantities in the evolution equations:
\bea
&& E = E_0 (1+z),\\
&&V_z = \frac{\omega_p^2 - m_a^2}{2 E} \propto \left\{
\bear{cc}
X_e (1+z)^2 & {\rm~for~}\omega_p^2 > m_a^2 \\
(1+z)^{-1} & {\rm~for~}\omega_p^2 < m_a^2, \\
\eear
\right.\\
&&V_x = g_a B \propto (1+z)^2, \\
&&D_0 \propto \left\{
\bear{cc}
X_e (1+z)^2 & {\rm~for~} E > m_e \\
X_e (1+z)^3 & {\rm~for~} E < m_e,
\eear
\right. \\
&&\ell_B^{-1} \propto (1+z), \\
&& H \propto \left\{
\bear{cc}
(1+z)^2 & {\rm~for~} z \gtrsim 3400 \\
(1+z)^\frac{3}{2} & {\rm~for~} z \lesssim 3400,
\eear
\right. .
\eea
In Fig.~\ref{fig:time} we show the evolution of these quantities as a function of the redshift. 
As one can see from the figure, some of them cross each other in the evolution of the Universe,
which will be important for evaluating the conversion rate. For later use we define the following redshift;
\bea
&&z=z_{\rm dec}  {\rm~~at~~}D_0 = H,\\
&&z=z_{\rm res}   {\rm~~at~~} \omega_p^2 = m_a^2.
\eea
Note that it is possible the parameters cross each other several times because $X_e$ changes between $1$ and $2 \times 10^{-4}$
at the recombination and reionization epoch. For simplicity we consider the case where the cross-over takes place
only once in the following analytical estimate, but such effects are taken into account in our numerical calculations. 
See Fig.~\ref{fig:Vz} for the evolution of $V_z$. One can see that the resonance takes place three times 
for $m_a = 10^{-13}$\,eV and $E_0/T_0 = 10^3$, while it takes place only once in the other cases. 

Solving Eq.~(\ref{Peq}) and deriving the conversion probability of the axion into photon are involved
because of the Hubble expansion which causes a resonant conversion at $V_z \sim 0$.
Below we derive the conversion probability in the off-resonant regime and resonant regime separately.

\subsection{Off resonant regime}

Let us first consider a simplified case without photon scatterings in a 
constant magnetic field ${\vec B}$, namely, $D_0 = \ell_B^{-1} = 0$.
From Eq.~(\ref{Peq}) one can see that the vector $\bf P$ rotates around the vector $\bf V$ with a precession frequency
$V \equiv \sqrt{V_x^2+V_z^2}$. 
In this case we can solve Eq.~(\ref{Peq}) analytically to find 
\begin{equation}
	P_{a\to\gamma}(t) =\frac{1}{2}(P_0(t)+P_z(t)) = \sin^2(2\theta) \sin^2\left( \frac{V t}{2} \right),
	\label{P_vac}
\end{equation}
where
\begin{equation}
	\sin^2(2\theta) = \frac{V_x^2}{V_x^2+V_z^2} = \frac{(2Eg_aB)^2}{(2Eg_aB)^2+(\omega_p^2-m_a^2)^2}.
	\label{sin2theta}
\end{equation}
In the expanding Universe, the energy as well as the magnetic field are redshifted. 
From the redshift-dependence of $V_x$ and $V_z$, one can see that $\theta$ remains almost constant 
for $\omega_p^2 > m_a^2$, while it scales as $(1+z)^3$ for $\omega_p^2 < m_a^2$.
Hereafter we consider the case of $|\theta| \simeq |V_x /2V_z| \ll 1$ except for the small time interval around the resonant regime 
($\omega_p^2 = m_a^2$), since otherwise the axion-photon mixing would be  too large to be consistent with observations.

Here it is should be noticed that the precession frequency $V$ is much larger than the Hubble 
parameter as well as the inverse of the coherent length $(\ell_B)$ of the magnetic field as long as $E_0/T_0 \lesssim 10^6$.
Thus the oscillations are averaged over the time of passing one domain of the coherent magnetic field.
If the axion passes through $N$ such regions with random orientation of the magnetic field with same strength, 
the conversion probability should be multiplied with the factor $N$~\cite{Grossman:2002by,Mirizzi:2005ng,Mirizzi:2009aj}
in the approximation that the change of the magnetic field among each domain is sudden.
In one Hubble time, $N \sim {\rm max}\{1, 1/(\ell_B H)\}$. As we shall see below, this effect will be automatically taken into account
in the presence of the damping term, $D = D_0 + \ell_B^{-1}$. 

\hspace{3mm}

Now we turn on the damping term. As mentioned before, the photons produced by the mixing will dissipate
into plasma at the rate of $D_0$.  If $D_0 > H$, those photons disappear in one Hubble time. 
On the other hand, if $D_0 < H$, the produced photons will free stream, thus contributing to the diffuse 
X-ray or $\gamma$-ray background or reionization at later times.  The conversion rate  depends on 
whether $D$ is larger or smaller than $V$. 
In the following we consider the following three cases; 
(i) $D_0 > V > {\rm max}[H, \ell_B^{-1}]$; (ii) $ V > D_0 > {\rm max}[H, \ell_B^{-1}]$ or  $V > \ell_B^{-1} > D_0 > H$;
(iii) $V > \ell_B^{-1} >  H > D_0$.  Note that, for the parameters of our interest, the Hubble parameter
is always smaller than $\ell_B^{-1}$, when $D_0 \approx \ell_B^{-1}$. 

In the case (i),  there is no effect of the coherent length of the  magnetic fields, i.e. $D \approx D_0$.
The polarization vector $\bf P$ sticks to the $z$ axis in this case, as there is no time for $P_x$ and $P_y$ to
evolve due to the large damping term.
Then $P_z$ evolves as
\begin{equation}
	\dot P_z \simeq -\frac{V_x^2}{D}P_z,
	\label{dtPz_D}
\end{equation}
where the dot represents the derivative with respect to time. 
Thus  $|{\bf P}| \approx |P_z|$ is reduced by
\begin{equation}
	|\Delta P_z| \sim \frac{V_x^2}{D H} = \frac{g_a^2 B^2}{DH}
	\label{DPz_D>k}
\end{equation}
in one Hubble time, where we have used a fact that $|P_z|$ is initially equal to $1$
and the conversion rate into photons is much smaller than unity. 
Note that $P_0$ changes by the same amount, $\Delta P_0 \simeq -\Delta P_z$, as long as $D_0 \gg H$, which is 
satisfied in the case (i).
Thus $P_0 + P_z$ remains zero, which implies that photons are scattered away soon after the production.
The injected photon energy density in one Hubble time is given by
\begin{equation}
\frac{\Delta\rho_\gamma}{s} \sim |\Delta P_z (t)| \frac{\rho_a}{s} \sim  \frac{g_a^2 B^2}{DH}  \lrf{\rho_a}{s}
~~~{\rm for~case~(i)}.
\end{equation}
Numerically, we have
\begin{equation}
	|\Delta P_z| \sim \frac{g_a^2 B^2}{DH} \sim 2\times 10^{-5}
	\left( \frac{g_a B_0}{10^{-16}\,{\rm GeV}^{-1}{\rm nG}} \right)^2 \left(\frac{10^4}{1+z}\right),
	\label{DP_case1}
\end{equation}
in the radiation dominated era for $E \lesssim m_e$.
In the matter dominated era, the most right hand side of (\ref{DP_case1}) should be multiplied by $\sqrt{(1+z)/3400}$.

We next consider the case (ii), in which $D_0$ is smaller than $V$, but is still larger than $H$. 
Note that $D_0$ becomes smaller than $V$ at a certain point, because
it decreases as $(1+z)^3$ while $V$ decreases as $(1+z)^2$ at early times  
$(\omega_p^2 > m_a^2)$ and increases as $(1+z)^{-1}$ at late times $(\omega_p^2 < m_a^2)$  (see Fig.~\ref{fig:time}). 
Also, $D_0$ will become smaller than $\ell_B^{-1}$ as the latter decreases more slowly, $\ell_B^{-1} 
\propto (1+z)$. 
In any case, the produced photons will dissipate into the plasma because the photon scatterings are
still frequent, i.e., $D_0 > H$. 

In the case (ii), the $\bf P$ rotates around the $\bf V$ with a frequency $V$,
while its amplitude gradually decreases due to the small damping.
Then the oscillation-averaged $\langle P_z\rangle$ evolves as
\begin{equation}
	\frac{d}{dt}\langle P_z\rangle \simeq -\frac{V_x^2D}{V_z^2}\langle P_z\rangle.
	\label{dtPz_D2}
\end{equation}
Thus the decrease of $|P_z|$ in one Hubble time is 
\begin{equation}
	|\Delta P_z| \sim \frac{V_x^2D}{V_z^2H}.
	\label{DPz_D<k}
\end{equation}
The injected photon energy density in one Hubble time is given by
\begin{equation}
\frac{\Delta\rho_\gamma}{s}  \sim  \frac{4E ^2g_a^2 B^2 D}{{\rm max}[\omega_p^4,m_a^4]H}  \lrf{\rho_a}{s}
~~~{\rm for~case~(ii)}.
\end{equation}
This formula contains a factor $D/H$, which is equal to $H^{-1}/\ell_B$ for $\ell_B^{-1} > D_0$. 
This factor represents the number of the cells the axion passes through in one Hubble time,
as discussed at the beginning of this section. 
Numerically, we have
\begin{equation}
	|\Delta P_z| \sim \frac{V_x^2D}{V_z^2H} \sim 8\times 10^{-21}\left( \frac{E_0}{T_0} \right)^2
	\left( \frac{g_a B_0}{10^{-16}\,{\rm GeV}^{-1}{\rm nG}} \right)^2 \left(\frac{1+z}{10^4}\right),
	\label{DP_case2}
\end{equation}
for $E \lesssim m_e$ and $D_0 > \ell_B^{-1}$ if $m_a \ll \omega_p$ in the radiation dominated era.
In the matter dominated era, the most right hand side of (\ref{DP_case2}) should be multiplied by $\sqrt{(1+z)/3400}$.

Lastly let us consider the case (iii). In this case the decrease of $|P_z|$ in one Hubble time is
similarly given by (\ref{DPz_D<k}). What is different is that the evolution of $P_0$ no longer follows
$-P_z$, and in general, $P_0 + P_z \ne 0$. This implies that the produced photons do not dissipate 
into plasma, but free stream. The free-streaming photons are accumulated as they are produced
by the mixing, and so, it is important to evaluate the timing when most of the free-stream photons are generated. 
The conversion rate in one Hubble time is given by~\footnote{
In the case of free-streaming photons, it is $|\Delta P_z|/2$  that represents the conversion rate, where the factor
$1/2$ arises from the fact that $P_0$ no longer follows the evolution of $P_z$. 
}
\beq
\frac{|\Delta P_z|}{2} \simeq \frac{V_x^2 \ell_B^{-1} }{2 V_z^2H} \propto \left\{
\bear{cc}
X_e^{-2} (1+z)^{- \frac{1}{2}}&  {\rm~for~}\omega_p^2 > m_a^2 \\
(1+z)^{\frac{11}{2}} & {\rm~for~}\omega_p^2 < m_a^2 \\
\eear
\right.,
\eeq
where we have used the fact that the Universe is matter-dominated for $z < z_{\rm dec}$.
Thus, the conversion rate increased until the resonance at $z = z_{\rm res}$, and then decreased
afterwards. 
The density of such free-streaming photons is determined at $z = {\rm min}[ z_{\rm dec}, z_{\rm res}]$:
\begin{equation}
\frac{\rho_\gamma}{s} \sim
 \lrf{2E ^2g_a^2 B^2}{m_a^4 H \ell_B}_{z= {\rm min}[z_{\rm dec}, z_{\rm res}]}  \lrf{\rho_a}{s}
{\rm~~~for~ case~(iii)}.
\end{equation}
If the resonance does not occur by the present time, it is evaluated at $z=0$.
Thus produced frree-streaming photons contribute  either to diffuse X-ray or $\gamma$-ray background
or to the energy injection at the reionization epoch if $E > 13.6$\,eV.

\subsection{Resonant regime}

Next we consider the conversion of axions into photons in the resonant regime, 
$|\omega_p^2 - m_a^2| \lesssim m_a^2$. To see what happens in this case,
let us first assume that the photon scattering are negligible and the magnetic field
is constant in space, namely, $D_0 = \ell_B^{-1} = 0$. Then,
when $\omega_p^2$ becomes equal to $m_a^2$ at $z = z_{\rm res}$, $V_z$ vanishes and the mixing 
angle becomes maximal,  $\theta \sim \pi/4$.  Before the mixing angle become maximal, however,
the polarization vector $\bf P$ ceases to follow the time-dependent $\bf V$ which changes so quickly at the resonance.
Below we  study the evolution of the axion-photon system 
around one Hubble time in the resonant region : $\Delta m^2 \equiv |\omega_p^2-m_a^2| \lesssim m_a^2$.

First, neglecting the damping term, one can see that the mixing angle is close to maximal for
$\Delta m^2 \lesssim (\Delta m^2)_{\rm max}  \equiv 2Eg_a B$.
Let us define the adiabaticity parameter $\alpha_V \equiv |\dot V / V^2| \simeq |\dot V_z/V_z^2|$,
which is given by $\alpha_V \simeq 6 H E m_a^2/(\Delta m^2)^2$ in the resonant regime. 
The polarization vector $\bf P$ precesses around $\bf V$ and follows its evolution 
while $\alpha_V  \lesssim 1$, or equivalently,
\begin{equation}
	\Delta m^2 \gtrsim (\Delta m^2)_{\rm adi} \equiv \sqrt{6 HE m_a^2}.
\end{equation}
Here and in what follows we assume $(\Delta m^2)_{\rm max} < (\Delta m^2)_{\rm adi}$, since otherwise 
a significant fraction of axions would be converted into photons in contradiction with observations,
 unless the initial axion density is negligibly small. 
When $\Delta m^2$ becomes equal to $\Delta m^2_{\rm adi}$, $\bf P$ ceases to follow $\bf V$. 
Thus, the 
the conversion rate is maximized at $\alpha_V \sim 1$ or equivalently $\Delta m^2 \sim (\Delta m^2)_{\rm adi}$, and given by
\begin{equation}
	(\Delta P_z)_{\rm res} \simeq 
\left.	\frac{2Eg_a^2 B^2}{3Hm_a^2}\right|_{z_{\rm res}},
	\label{Pz_res}
\end{equation}
where all the parameters here are evaluated at the resonance.
In the absence of the photon scatterings, the change in $P_z/2$ results in the increase of the photon number density,
and this fraction of axions are converted to free-streaming photons. Thus we obtain
\begin{equation}
	\frac{\rho_\gamma}{s} \simeq \frac{\left.\Delta P_z \right|_{z_{\rm res}}}{2} \frac{\rho_a}{s}
	\sim  \left.\frac{E g_a^2 B^2}{3 H m_a^2}\right|_{z_{\rm res}}\frac{\rho_a}{s}~~~~{\rm for~} z < z_{\rm res}
\end{equation}
This agrees with the result in Refs.~\cite{Mirizzi:2009iz,Mirizzi:2009nq} up to a factor of order unity.

Let us now turn on the damping terms, $D_0$ and $D$. If $D$ is greater than $m_a^2/2E$ at $z = z_{\rm res}$,
the effective mixing angle remains suppressed by the damping term, and
there is no resonant production. We therefore focus on the case where $D$ is smaller 
than $V$ before the system enters the resonant regime. In this case, 
 $D$ becomes greater than $V (\sim |V_z|)$ only in the vicinity of the resonant point,
$\Delta m^2 \lesssim (\Delta m^2)_D \equiv 2ED$. Let us consider the effect of the damping term
on the photon production and its dissipation. 
 If $(\Delta m^2)_D < (\Delta m^2)_{\rm adi}$, the damping term does not have significant effect
and $(\Delta P_z)_{\rm res}$ after the resonant region is given by Eq.~(\ref{Pz_res}).
Then the $P_0$ changes by $(\Delta P_0)_{\rm res} \sim -D_0 \Delta t (\Delta P_z)_{\rm res}$ (see Eq.~(\ref{Peq})),
where $\Delta t$ is the time interval during which the adiabaticity is broken, and is given by $D\Delta t \sim (\Delta m^2)_D / (\Delta m^2)_{\rm adi} \lesssim 1$.  This implies that a fraction $D_0 \Delta t (\lesssim 1)$ of the produced photons are scattered during
the time interval $\Delta t$. The rest of ``free-streaming'' photons dissipate into the plasma in one Hubble time around the resonance
if $D_0 > H$, while they remain free-streaming if $D_0 < H$. 
%
Therefore, we obtain
\begin{equation}
	\left.\frac{\Delta\rho_\gamma}{s}\right|_{z_{\rm res}} \simeq \frac{(\Delta P_z)_{\rm res}}{2} \left.\frac{\rho_a}{s}\right|_{z_{\rm res}}
	\sim  \left.\frac{E g_a^2 B^2}{3 H m_a^2}\right|_{z_{\rm res}} \left.\frac{\rho_a}{s}\right|_{z_{\rm res}},
\end{equation}
for $D \gg H$ at $z=z_{\rm res}$, and
\begin{equation}
	\left.\frac{\rho_\gamma}{s}\right|_{z < z_{\rm res}} \simeq \frac{(\Delta P_z)_{\rm res}}{2} \frac{\rho_a}{s}
	\sim  \left.\frac{E g_a^2 B^2}{3 H m_a^2}\right|_{z_{\rm res}}\frac{\rho_a}{s},
\end{equation}
for $D \ll H$ at $z<z_{\rm res}$.
Note that $(\Delta P_z)_{\rm res}$ is larger than the off-resonant value at $z\sim z_{\rm res}$ (Eq.~(\ref{DP_case2})) by a factor 
$\sim m_a^2/2ED$.

On the other hand, if $(\Delta m^2)_D > (\Delta m^2)_{\rm adi}$, the evolution of $P_z$ is governed by the Eq.~(\ref{dtPz_D})
for the time interval $\Delta t \sim 2ED/(3Hm_a^2)$.
Thus during the damping regime, $P_z$ changes with an amount
\begin{equation}
	(\Delta P_z)_{\rm res} \sim \frac{V_x^2}{D}\Delta t \sim \left.\frac{2E g_a^2 B^2}{3 H m_a^2}\right|_{z_{\rm res}}.
\end{equation}
Again, all quantities here are evaluated at the resonant region. This is same expression as (\ref{Pz_res}).
This also agrees with the result in Refs.~\cite{Jaeckel:2008fi,Mirizzi:2009iz,Mirizzi:2009nq}.
Since $D \Delta t \gtrsim 1$ in this case, we have $(\Delta P_0)_{\rm res} \sim -(\Delta P_z)_{\rm res}$ and hence the injected photon energy density is given by
\begin{equation}
	\left.\frac{\Delta\rho_\gamma}{s}\right|_{z_{\rm res}} \sim -(\Delta P_0)_{\rm res} \left.\frac{\rho_a}{s}\right|_{z_{\rm res}}
	\sim  \left.\frac{2E g_a^2 B^2}{3 H m_a^2}\right|_{z_{\rm res}} \left.\frac{\rho_a}{s}\right|_{z_{\rm res}}.
\end{equation}
Numerically, $(\Delta P_0)_{\rm res}$ is evaluated as
\begin{equation}
	(\Delta P_0)_{\rm res} = \left.\frac{2E g_a^2 B^2}{3 H m_a^2}\right|_{z_{\rm res}}
	\sim 4\times 10^{-13}\left( \frac{E_0}{T_0} \right)\left( \frac{g_a B_0}{10^{-16}\,{\rm GeV}^{-1}{\rm nG}} \right)^2
	\left( \frac{10^{-8}\,{\rm eV}}{m_a} \right)^2\left( \frac{1+z_{\rm res}}{10^4} \right)^3,
	\label{DPres_num}
\end{equation}
if the resonance happens at the radiation dominated era $(z_{\rm res} \gtrsim 3400)$.
Note that $z_{\rm res} ^3 \propto m_a^2$ and hence this expression does not depend on $m_a$.
If $z_{\rm res} \lesssim 3400$, the most right hand side of (\ref{DPres_num}) should be multiplied by the factor $\sqrt{(1+z_{\rm res})/3400}$.

\section{Constraints on axionic dark radiation}
\label{sec:const}

\begin{figure}
\begin{center}
\includegraphics[scale=1.2]{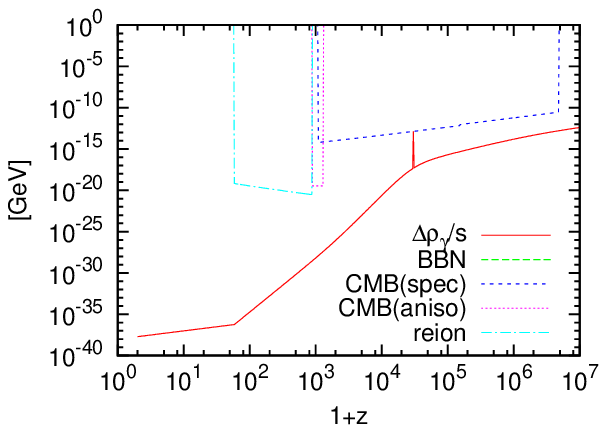} 
\includegraphics[scale=1.2]{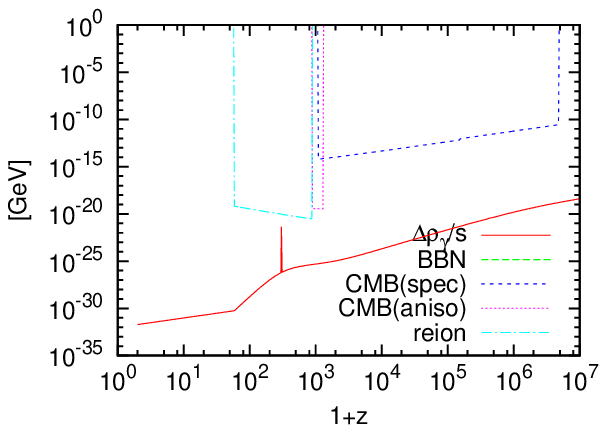} 
\includegraphics[scale=1.2]{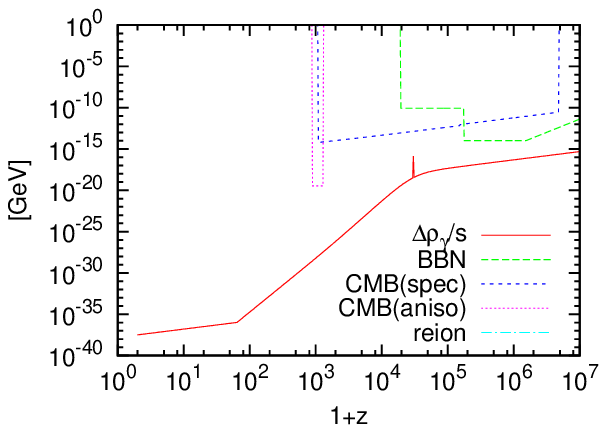} 
\includegraphics[scale=1.2]{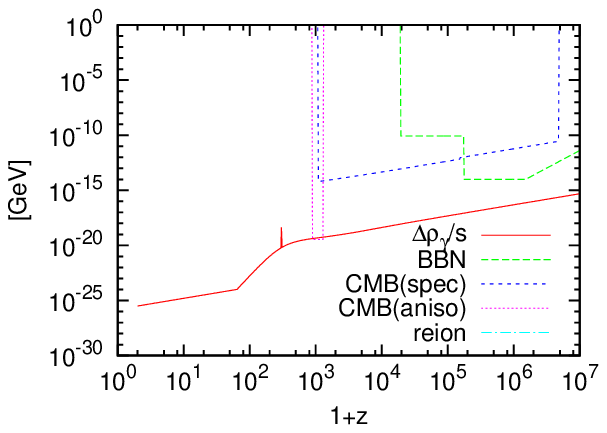} 
\caption{
	$\Delta \rho_\gamma/s$ as a function of redshift $z$ for
	$(E_0/T_0, m_a, g_a B_0) = (10^3, 10^{-7}\,{\rm eV}, 10^{-13}\,{\rm GeV^{-1} nG})$ (top left),
	$(10^3, 10^{-10}\,{\rm eV}, 10^{-16}\,{\rm GeV^{-1} nG})$ (top right),
	$(10^6, 10^{-7}\,{\rm eV}, 10^{-16}\,{\rm GeV^{-1} nG})$ (bottom left) and
	$(10^6, 10^{-10}\,{\rm eV}, 10^{-16}\,{\rm GeV^{-1} nG})$ (bottom right).
	Together shown are upper bounds from various cosmological observations (see text).
}
\label{fig:Deltarho}
\end{center}
\end{figure}

\begin{figure}
\begin{center}
\includegraphics[scale=0.9]{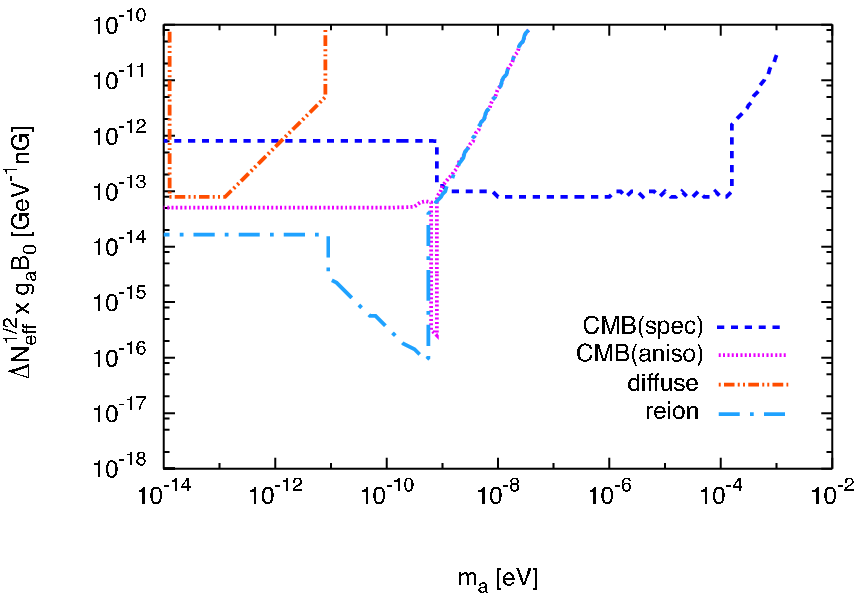} 
\includegraphics[scale=0.9]{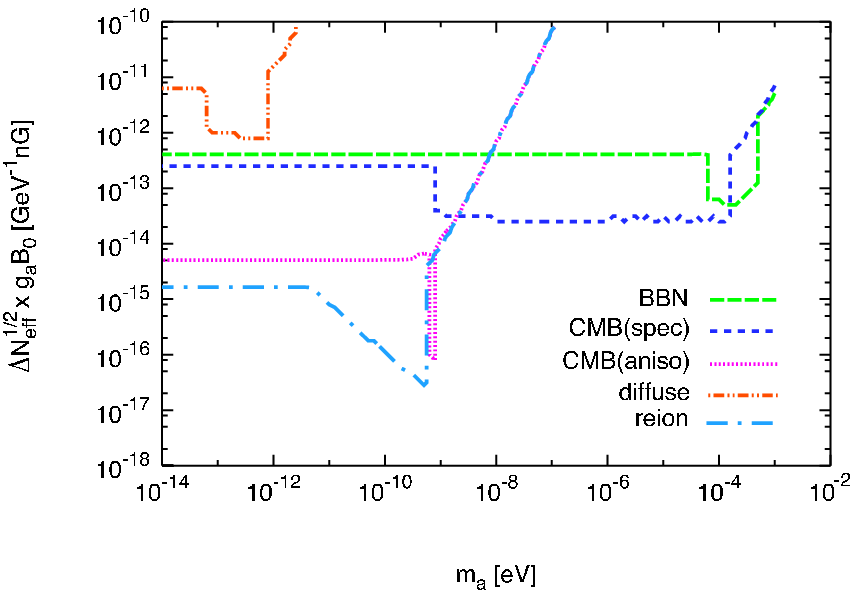} 
\includegraphics[scale=0.9]{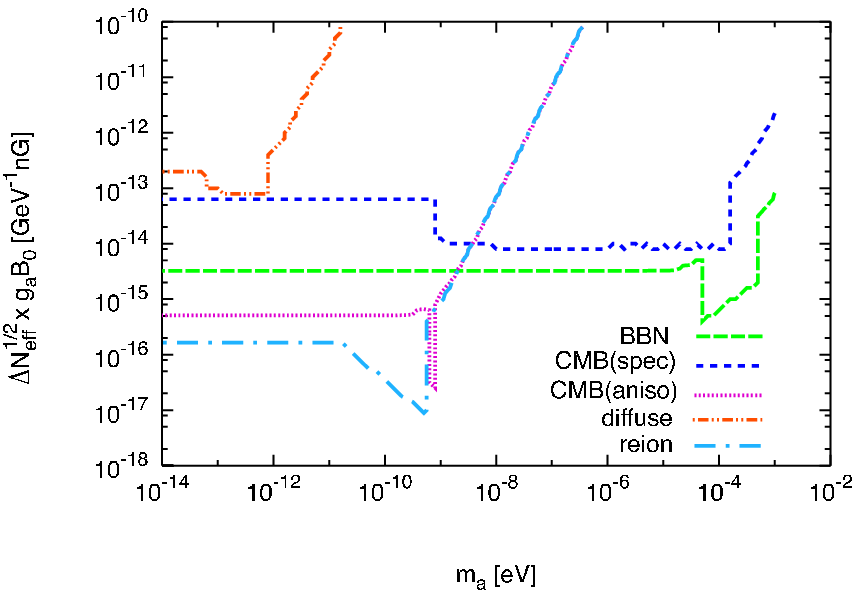} 
\includegraphics[scale=0.9]{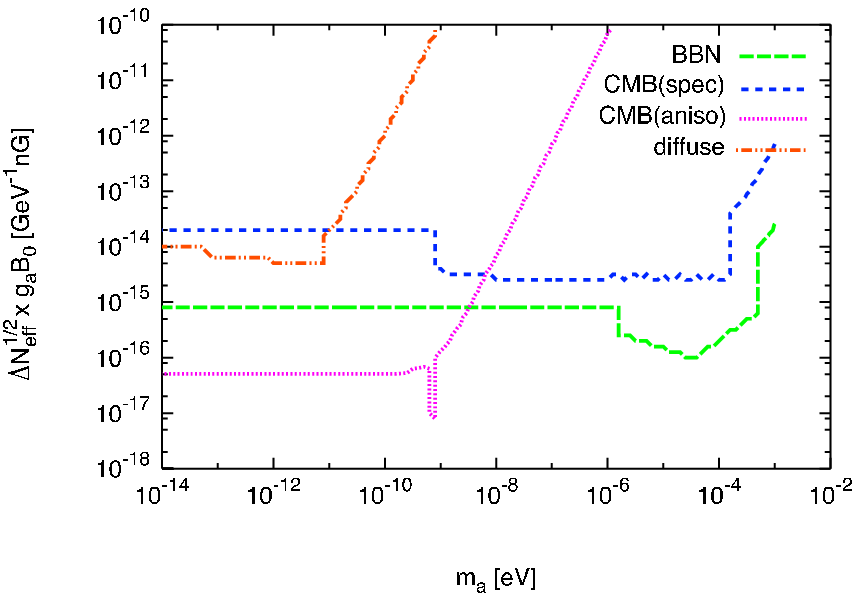} 
\caption{
	Constraints on ($m_a, \Delta N_{\rm eff}^{1/2} \times g_a B_0$) plane for
	$E_0/T_0 = 10^3$ (top left),
	$E_0/T_0 = 10^4$ (top right),
	$E_0/T_0 = 10^5$ (bottom left),
	 and $E_0/T_0 = 10^6$ (bottom right).
}
\label{fig:const}
\end{center}
\end{figure}

As shown in Refs.~\cite{Higaki:2012ba,Cicoli:2012aq,Higaki:2012ar,Higaki:2013lra,Angus:2013zfa},
the branching ratio of the modulus decay into its light axion partner is generically substantial if the modulus is stabilized by
SUSY breaking effects. Moreover, the moduli can also decay into the axion component of other moduli, which may support the SM cycle.

The abundance of relativistic axions is parameterized by the effective number of neutrino species, $\Delta N_{\rm eff}$, given by
\begin{equation}
	\Delta N_{\rm eff} = \frac{43}{7}\left( \frac{10.75}{g_{*s}(T_d)} \right)^{1/3}\frac{B_a}{1-B_a},
\end{equation}
where $T_d$ is the modulus decay temperature and $B_a$ the modulus branching ratio into the axion pair.
Here it is assumed that modulus dominates the Universe at the decay.
The Planck constraint reads $\Delta N_{\rm eff} < 0.84$ at the 95\% C.L.~\cite{Ade:2013lta}.
Using this, the ratio of the axion energy density to the entropy density after the $e^+e^-$ annihilation is given by
\begin{equation}
	\frac{\rho_a}{s} = \Delta N_{\rm eff} \frac{7}{8}\left( \frac{4}{11} \right)^{1/3} \frac{\rho_\gamma}{s}
	\simeq 0.24\,\Delta N_{\rm eff}\,T_\gamma,
\end{equation}
where $T_\gamma$ is the photon temperature.

As shown in the previous section, a part of the relativistic axions is converted into the photon under the primordial magnetic field
and it  causes various cosmological effects depending on the redshift and the axion energy.
The cosmological constraints on the presence of such high-energy photon injection were summarized in Ref.~\cite{Kawasaki:2007mk},
which, in the present case, reads
\begin{equation}
	\frac{\Delta\rho_\gamma}{s} \lesssim \left( \frac{\rho_\gamma}{s} \right)_{\rm bound},
\end{equation}
where the right hand side means the upper bound on the injected photon energy density in one Hubble time.
The bound includes the followings (see Ref.~\cite{Kawasaki:2007mk} for further details).

\begin{itemize}
\item BBN :
Photodissociation of light elements are caused by the additional photons.
This does not occur for $E<4.5$\,MeV, below the threshold energy of the D destruction.
For $E>4.5$\,MeV, this gives tight constraint at $t \gtrsim 10^{7}$\,sec.

\item Spectral distortion of CMB :
Additional photons are not thermalized for $t \gtrsim 10^{6}$\,sec and hence may distort the blackbody spectrum of the CMB.

\item Recombination :
Photons with energies $E > 13.6$\,eV injected at the recombination epoch $z_{\rm rec}\sim 1090$ may affect the ionization fraction of the hydrogen atom, which results in the change in the power spectrum of the CMB anisotropy.
This leads to severe constraint : $(\rho_\gamma/s)_{\rm bound} \sim 4\times 10^{-20}$\,GeV at around $z \sim 1090$.

\item Diffuse photon background :
Photons with keV -- GeV ranges after the recombination are transparent and may be observed as diffuse X\,$(\gamma)$-ray background. For low energy photons with $E < 13.6$\,eV also contribute to diffuse background photons.
We also assumed that photons with all energies of our interest are transparent after the reionization $z < 11.4$.

\item Reionization :
Photons with energies between $13.6$\,eV and $\sim$ keV after the recombination but before the reionization
contribute as extra ionization sources for the neutral hydrogen,
which may lead to too large optical depth to the last scattering surface as indicated from the measurement of the CMB anisotropy.

\end{itemize}

Figs.~\ref{fig:Deltarho} show $\Delta \rho_\gamma/s$ as a function of redshift $z$ for
$(E_0/T_0, m_a, g_a B_0) = (10^3, 10^{-7}\,{\rm eV}, 10^{-13}\,{\rm GeV^{-1} nG})$ (top left),
$(10^3, 10^{-10}\,{\rm eV}, 10^{-16}\,{\rm GeV^{-1} nG})$ (top right),
$(10^6, 10^{-7}\,{\rm eV}, 10^{-16}\,{\rm GeV^{-1} nG})$ (bottom left) and
$(10^6, 10^{-10}\,{\rm eV}, 10^{-16}\,{\rm GeV^{-1} nG})$ (bottom right).
In these figures we have fixed $\Delta N_{\rm eff} = 0.5$.
Together shown are upper bounds from various cosmological observations.
As one changes $g_a B_0$, the solid (red) line goes up and down. The upper bound on $g_a B_0$
is obtained in each case so that the solid (red) line touches the constraint lines.

Fig.~\ref{fig:const} shows the resulting constraints on ($m_a, \Delta N_{\rm eff}^{1/2} g_a B_0$) plane for 
$E_0/T_0 = 10^3$ (top left), $E_0/T_0 = 10^4$ (top right), $E_0/T_0 = 10^5$ (bottom left) and $E_0/T_0 = 10^6$ (bottom right).
Typically the conversion rate is larger for higher redshift, hence the constraint from diffuse background photons is not so stringent.
Instead, the CMB anisotropy constrains the photon injection around the recombination epoch and it gives tight bound.
For larger $E_0$, the converted photon energy is sufficiently high to destroy light elements and the BBN constraint becomes important.
It is seen that the combination $g_a B_0$ is tightly constrained for a wide range of the axion mass  shown in the figure.
Note that we have adopted the present coherent length of the magnetic field to be $1$\,Mpc.
If it is smaller, the constraint becomes severer as axions pass through a larger number of the magnetic cells,
leading to an enhancement of the conversion rate.

\section{Discussion and Conclusions}
\label{sec:disc}

In this paper we have studied cosmological effects of the axionic dark radiation in the presence of primordial magnetic field.
We have derived constraints on the axion-photon coupling and the strength of the primordial magnetic field.
If future observations confirm the primordial magnetic field, it will give robust constraints on the properties of axion dark radiation. 
Since a substantial amount of axionic dark radiation is often produced in the modulus/saxion decay in SUSY axion models or concrete compactification models of string theory, it also gives an important constraint on the high-energy theory.
On the other hand, as shown recently in Ref.~\cite{Conlon:2013isa}, the axion helioscope may be able to detect relativistic axion background if $g_a$ is relatively large. Then it will give tight constraints on the primordial magnetic field.

Let us comment on the case of the QCD axion.
In fact, our constraint is so severe that the a part of the parameters for the QCD axion can be excluded
as a dominant dark radiation if the magnetic field has a primordial origin.
In the case of QCD axion, the axion-photon coupling $g_a$ is related to the axion decay contant as
\beq
g_a \simeq C \frac{\alpha}{\pi} \frac{1}{f_a},
\eeq
where $C \approx -0.97$ for the KSVZ axion and $C \approx 0.36$ for the DFSZ axion.
The axion mass is given by
\beq
m_a \simeq 6\times 10^{-5} {\rm \,eV} \lrf{10^{11}\,{\rm GeV}}{f_a} \sim (3 - 7) \times 10^{-5}\,{\rm eV} \lrf{g_a}{10^{-14}\,{\rm GeV}^{-1}}.
\eeq
One can see from Fig.~\ref{fig:const} that $(m_a, g_a) \approx (10^{-5} \,{\rm eV}, 10^{-14} {\rm\,GeV}^{-1})$ 
is exlucded for $E_0/T_0 = 10^5$ or $10^6$, $B_0 \approx 1$\,nG and 
$\Delta N_{\rm eff} \simeq 0.5$. For smaller $E_0/T_0$, $B_0$, and $\Delta N_{\rm eff}$, there is a room for the QCD
axion to be the dominant component of dark radiation.
Note that the QCD axion can naturally explain dark matter for $f_a = 10^{11 - 12}$\,GeV in the
absence of entropy production after the QCD phase transition. In our scenario, there may be a large entropy production
by the modulus decay, in which case the axion decay constant as large as $\GEV{15}$ is allowed without fine-tuning of the 
initial misalignment angle.

Note also that, although we have focused on the cosmological effects of the axion-photon conversion,
the Galactic magnetic field also converts the axionic dark radiation into photons.
From Eq.~(\ref{sin2theta}), the conversion probability is given by
\begin{equation}
	P_{a\to \gamma} \sim 10^{-13}\left( \frac{E_0}{1\,{\rm eV}} \right)^2\left( \frac{g_a}{10^{-10}\,{\rm GeV}^{-1}} \right)^2
	\left( \frac{B_{\rm Gal}}{1\,\mu{\rm G}} \right)^2
	\left( \frac{10^{-10}\,{\rm eV}}{m_a} \right)^4,
	\label{conv_gal}
\end{equation}
where $B_{\rm Gal}$ is the typical magnetic field strength in the Galaxy.\footnote{
	The probability (\ref{conv_gal}) may depend on the detailed structure of the Galactic magnetic field.
	Here we have neglected it.
}
The oscillation length is much shorter than the typical coherent scale of the Galactic magnetic field ($\sim 1$\,pc).
The conversion probability is saturated at $m_a \sim 10^{-11}\,{\rm eV}$, below which the plasma frequency becomes important.
Thus typically the conversion rate is small,
but it may be more important than the cosmological one depending on the value of $B_0$.

Some comments are in order.
Since the modulus in general dominates the Universe, the pre-existing primordial magnetic field is diluted accordingly.
Therefore we may need efficient mechanism for creating the magnetic field to explain observations~\cite{Durrer:2013pga}.
If the currently observed magnetic field is not of the primordial origin, but produced during the structure formation, there may be
effectively no magnetic field in the early Universe. In this case,
most of the constraints derived in this paper are not applied. It should be noted however that the generation of the magnetic
field and its subsequent evolution are complicated issues, and it is even possible that sufficiently large magnetic fields are
produced as a result of amplifications due to the turbulent small-scale dynamo~\cite{Wagstaff:2013yna}. If this is the case,
our constraints will provide extremely tight constraints on the amount of axion dark radiation and its properties.

\section*{Acknowledgments}

We would like to thank the YITP at Kyoto University for the hospitality
during the YITP workshop YITP-W-12-21 on ``LHC vs Beyond the Standard Model'',
where the present work started. 
This work was supported by 
the Grant-in-Aid for Scientific Research on Innovative Areas
(No.~24111702 [FT], No.~21111006 [KN and FT], and No.~23104008 [FT]),  
Scientific Research (A) (No.~22244030 [KN and FT] and No.~21244033 [FT]), 
JSPS Grant-in-Aid for Young Scientists (B) (No. 24740135 [FT] and No. 25800169 [TH]),
and Inoue Foundation for Science [FT].


\end{document}